\documentclass[english]{article}
\usepackage[T1]{fontenc}
\usepackage[latin1]{inputenc}
\usepackage{geometry}
\makeatletter

\providecommand{\tabularnewline}{\\}

\usepackage{babel}
\makeatother
\begin{document}

\title{Accurate calculations of magnetic dipole and electric quadrupole
hyperfine coupling constants of 6$P_{3/2}$ state of Cesium atom :
A Relativistic Coupled Cluster approach}

\author{Rajat K Chaudhuri and Chiranjib Sur \\
\emph{NAPP Group, Indian Institute of Astrophysics, Bangalore - 560
034, INDIA}}

\date{(Received date, Accepted date)}

\maketitle
\begin{abstract}
We report the magnetic dipole and electric quadrupole hyperfine coupling
constants of 6$P_{3/2}$ state of $^{133}$Cs(I=$\frac{7}{2})$ obtained
from the relativistic coupled cluster (RCC) method. To our knowledge,
no prior electric quadrupole hyperfine CC calculation is available
for $^{133}$Cs(6$P_{3/2}$) state. Our computed magnetic dipole ($A$)
and electric quadrupole ($B$) coupling constants are in excellent
agreement with the experiment. 

\textbf{PACS number(s).} : 31.15.Ar, 31.15.Dv, 31.25.Jf, 32.10.Fn
\end{abstract}
Theoretical determination of hyperfine coupling constant is one of
most stringent tests of atomic wave functions near the nucleus. Also
accurate prediction of hyperfine coupling constants and other related
properties (viz. transition energies) requires precise incorporation
of relativistic and higher correlation and relaxation effects as these
effects are strongly entangled. In particular, the hyperfine coupling
constant calculations are relevant to the parity non-conservation
study in atoms because the electro-weak interaction is also a short
range force.

The relativistic and dynamical electron correlation effects can be
incorporated in many-electron systems through a variety of many-body
methods. The coupled cluster method (CCM) has emerged as one of the
most powerful and effective tool for a high precision description
of electron correlations in many-electron systems. The CCM is an all-order
non-perturbative scheme, and therefore, the higher order electron
correlation effects can be incorporated more efficiently than using
the order-by-order diagrammatic many-body perturbation theory (MBPT).
The CC method is size-extensive, a property which has been found to
be crucial for an accurate determination of state energies, bond cleavage
energies and related spectroscopic constants. Since the order-by-order
MBPT expansion terms are directly related to the CC equations (as
the latter is an all-order version of the former scheme), the CC results
can be improved by adding the important omitted diagrams with the
aid of low order MBPT.

Precise measurements of hyperfine splittings in $6P_{3/2}$ state
of $^{133}$Cs have been carried out by Tanner and Weiman \cite{Tanner}.
The magnetic dipole and electric quadrupole hyperfine coupling constant
of $6P_{3/2}$ state have also been determined from this high precision
experiment. In this article, we report the magnetic dipole ($A$)
and electric quadrupole ($B$) hyperfine coupling constant of $6P_{3/2}$
state of Cesium atom obtained from the RCC method to access the performance
and accuracy of the RCC scheme. We reiterate that accurate computation
of hyperfine coupling constants requires precise incorporation of
electron correlations and accurate description of the wave functions
near the nucleus.

The relativistic hyperfine energy $W$ for states $J=L\pm\frac{1}{2}$
is given by \begin{equation}
W=A(I.J)+B\frac{6(I.J)^{2}+3(I.J)-2((I.I)(J.J)}{2I(I-1)2J(J-1)},\label{energy}\end{equation}
 where $I$ is the nuclear spin. $A$ and $B$ are, respectively,
the coefficients of magnetic dipole and electric quadrupole contribution
to the hyperfine structure. The radial matrix elements related to
$A$ and $B$ are available elsewhere \cite{Cheng}. {[}Note that
$B$=0 for $J=\frac{1}{2}$.{]}

\noindent We employ the straight forward extension of non-relativistic
open-shell coupled cluster theory to the relativistic regime by adopting
the no-virtual-pair approximation (NVPA) along with appropriate modification
of orbital form and potential terms \cite{Kaldor}. We begin with
Dirac-Coulomb Hamiltonian ($H_{D}$) expressed in normal order \begin{equation}
H_{D}=H_{N}-\langle0|H|0\rangle=\sum_{ij}\langle i|f|j\rangle\left\{ a_{i}^{\dagger}a_{j}\right\} +\frac{1}{4}\sum_{i,j,k,l}\langle ij||kl\rangle\left\{ a_{i}^{\dagger}a_{j}^{\dagger}a_{l}a_{k}\right\} ,\label{eq2}\end{equation}
 where \begin{equation}
\langle ij||kl\rangle=\langle ij|\frac{1}{r_{12}}|kl\rangle-\langle ij|\frac{1}{r_{12}}|lk\rangle.\label{eq3}\end{equation}

\noindent The valence universal Fock space open-shell coupled cluster
method is employed which begins with the decomposition of the full
many-electron Hilbert space of dimension $N$ into into a reference
space $\mathcal{M}_{0}$ of dimension $M\ll N$, defined by the projector
$P$, and its orthogonal complement $\mathcal{M}_{0}^{\perp}$ associated
with the projector $Q=1-P$. A valence universal wave operator $\Omega$
is then introduced which satisfies \begin{equation}
|\Psi_{i}\rangle=\Omega|\Psi_{i}^{(0)}\rangle,\;\;\; i=1,\ldots,M,\label{eq4}\end{equation}
 where $|\Psi_{i}^{(0)}\rangle$ and $|\Psi_{i}\rangle$ are the \emph{unperturbed}
and \emph{the exact} wave functions of the $i$th eigenstate of the
Hamiltonian, respectively. The wave operator $\Omega$, which formally
represents the mapping of the reference space $\mathcal{M}_{0}$ onto
the target space $\mathcal{M}$ spanned by the $M$ eigenstates $|\Psi_{i}\rangle$,
has the properties \begin{equation}
\Omega P=\Omega,\;\; P\Omega=P,\;\;\Omega^{2}=\Omega.\label{eq5}\end{equation}
 With the aid of the wave operator $\Omega$, the Schr\"{o}dinger
equation for the $M$ eigenstates of the Hamiltonian correlating with
the $M$-dimensional reference space, i.e. \begin{equation}
H|\Psi_{i}\rangle=E_{i}|\Psi_{i}\rangle,\;\;\; i=1,\ldots,M,\label{eq6}\end{equation}
 is transformed into a generalized Bloch equation, \begin{equation}
H\Omega P=\Omega H\Omega P=\Omega PH_{\textrm{eff}}P,\label{eq7}\end{equation}
 where $H_{\textrm{eff}}\equiv PH\Omega P$ is the effective Hamiltonian.
Once Eq. (\ref{eq6}) is solved for the wave operator $\Omega$, the
energies $E_{i}$, $i=1,\ldots,M$, are computed by diagonalizing
the effective Hamiltonian $H_{\textrm{eff}}$ in the $M$-dimensional
reference space $\mathcal{M}_{0}$. Following Lindgren's formulation
of open-shell CC \cite{Lindgren}, we express the valence universal
wave operator $\Omega$ as \begin{equation}
\Omega=\{\exp(S)\},\label{eq8}\end{equation}
 where $S$ is the excitation operator and curly brackets denote the
normal ordering.

Once $\Psi_{i}$ is known, the one-electron properties $\langle O\rangle$
are evaluated from the following expression \begin{equation}
\langle O\rangle_{fi}=\frac{\langle\Psi_{f}|O|\Psi_{i}\rangle}{\sqrt{\langle\Psi_{f}|\Psi_{f}\rangle\langle\Psi_{i}|\Psi_{i}\rangle}}.\label{eq8}\end{equation}

\noindent In the actual calculations, the ground and excited state
properties of Cs are computed using the finite basis set expansion
method (FBSE) \cite{FBSE} with a large basis set of Gaussian functions
of the form \begin{equation}
F_{i,k}(r)=r^{k}\cdot e^{-\alpha_{i}r^{2}},\label{eq9}\end{equation}
 with $k=0,1,\dots$ for $s$, $p$, $...$ type functions, respectively.
For the exponents, the even tempering condition \begin{equation}
\alpha_{i}=\alpha_{0}\beta^{i-1}\label{eq10}\end{equation}
 is applied. Here, $N$ is the number of basis functions for a specific
symmetry. The starting point of this calculation is the generation
of DF orbitals for Cs$^{+}$ which defines the (0,0) valence sector
of the Fock space. The ion is then correlated by CCSD and one-electron
is then added, following the Fock-space scheme \cite{Mukherjee}:

\[
\mathrm{Cs^{+}}(0,0)+e\longrightarrow\mathrm{Cs}(0,1)\]

\begin{table}

\caption{Comparison of hyperfine ($A$ and $B$) coupling (in MHz) and 6s$\rightarrow$6p$_{3/2}$
transition dipole matrix element of $^{133}$Cs atom with the experiment
and other correlated calculations.}

\begin{center}\begin{tabular}{lrrrr}
\hline 
&
 This work &
\multicolumn{2}{c}{Others}&
 Experiment \tabularnewline
&
&
&
&
\tabularnewline
\hline
\hline 
$A$(6$P_{3/2}$) &
 50.419 &
 49.785\cite{Blundell}&
 48.51\cite{Safronova}&
 50.275(3) \cite{Tanner}\tabularnewline
 $B$(6$P_{3/2}$) &
 -0.559 &
&
&
 -0.53(2) \cite{Tanner}\tabularnewline
 6$S_{1/2}\rightarrow6P_{3/2}$&
 6.372 &
 6.370\cite{Blundell}&
 6.304\cite{Safronova}&
 6.36 \cite{Shabanova} \tabularnewline
\hline
\end{tabular}\end{center}
\end{table}

Table 1 compares the magnetic dipole and electric quadrupole hyperfine
coupling matrix element of $^{133}$Cs(I=7/2) obtained from the CC
method with the experiment \cite{Tanner} and with linearized CC calculations
\cite{Blundell,Safronova}. We also report the $6s\rightarrow6p_{3/2}$
transition dipole matrix to demonstrate the consistency of this method.
Our computed magnetic dipole and electric quadrupole hyperfine coupling
matrix elements are quite accurate {[}off by 0.143 MHz for the magnetic
dipole and 0.02 MHz for electric quadrupole case{]}. The CC transition
dipole matrix element reported here is also in excellent agreement
with the experiment \cite{Shabanova}. Finally, we emphasize that
to our knowledge no prior electric quadrupole hyperfine CC calculations
are available for 6$P_{3/2}$ state of Cs atom.

\begin{verse}
\textbf{Acknowledgments} : One of the author (CS) acknowledge BRNS
for project no. 2002/37/12/BRNS.
\end{verse}


\begin{thebibliography}{1}
\bibitem{Tanner}C. E. Tanner and C. Weiman, Phys. Rev. A \textbf{38}, 1616 (1988).
\bibitem{Cheng}K.T. Cheng and W.J. Childs, Phys. Rev. A \textbf{31}, 2775 (1985).
\bibitem{Kaldor}E. Eliav, U. Kaldor, and Y. Ishikawa, Phys. Rev. A \textbf{49}, 1724
(1994).
\bibitem{Lindgren}I. Lindgren, Int. J. Quantum. Chem. \textbf{S12}, 33 (1978).
\bibitem{FBSE}R. K. Chaudhuri, P. K. Panda, and B. P. Das, Phys. Rev. A \textbf{59},
1187 (1999).
\bibitem{Mukherjee}I. Lindgren and D. Mukherjee , Physica Scripta \textbf{93}, 151 (1987).
\bibitem{Blundell}S. A. Blundell, W. R. Johnson, and J. Sapirstein, Phys. Rev. A \textbf{43},
3407 (1991).
\bibitem{Safronova}M. S. Safronova, W. R. Johnson, and A. Derevianko, Phys. Rev. A \textbf{60},
4476 (1999).
\bibitem{Shabanova}L. Shabanova, Y. Monakov, and A. Khlyustalov, Opt. Spectrosc. (USSR)
\textbf{47}, 3 (1979).\end{thebibliography}
\end{document}